\documentclass[lettersize,journal]{IEEEtran}
\usepackage{amsmath,amsfonts}
\usepackage{algorithmic}
\usepackage{algorithm}
\usepackage{array}
\usepackage[caption=false,font=normalsize,labelfont=sf,textfont=sf]{subfig}
\usepackage{textcomp}
\usepackage{stfloats}
\usepackage{url}
\usepackage{verbatim}
\usepackage{graphicx}
\usepackage{cite}
\usepackage{siunitx}
\usepackage{booktabs}
\usepackage[colorlinks = true,
            linkcolor = blue,
            urlcolor  = blue,
            citecolor =  green,
            anchorcolor = blue]{hyperref}

\begin{document}

\title{Separate Anything You Describe}

\author{Xubo Liu, Qiuqiang Kong, Yan Zhao, Haohe Liu, Yi Yuan, Yuzhuo Liu, \\Rui Xia, Yuxuan Wang, Mark D. Plumbley, Wenwu Wang\\ \url{https://audio-agi.github.io/Separate-Anything-You-Describe}
\thanks{Xubo Liu, Haohe Liu, Yi Yuan, Mark D. Plumbley, and Wenwu Wang are with the Centre for Vision, Speech and Signal Processing (CVSSP), University of Surrey, Guilford, UK. Email: \{xubo.liu, haohe.liu, yi.yuan, m.plumbley, w.wang\}@surrey.ac.uk.}
\thanks{Qiuqiang Kong, Yan Zhao, Yuzhuo Liu, Rui Xia, and Yuxuan Wang: are with the Speech, Audio \& Music Intelligence (SAMI) Group, ByteDance Inc. Email: \{kongqiuqiang, zhao.yan, liuyuzhuo.999, rui.xia, wangyuxuan.11\}@bytedance.com.}}

\maketitle

\begin{abstract}
Language-queried audio source separation (LASS) is a new paradigm for computational auditory scene analysis (CASA). LASS aims to separate a target sound from an audio mixture given a natural language query, which provides a natural and scalable interface for digital audio applications. Recent works on LASS, despite attaining promising separation performance on specific sources (e.g., musical instruments, limited classes of audio events), are unable to separate audio concepts in the open domain. In this work, we introduce AudioSep, a foundation model for open-domain audio source separation with natural language queries. We train AudioSep on large-scale multimodal datasets and extensively evaluate its capabilities on numerous tasks including audio event separation, musical instrument separation, and speech enhancement. AudioSep demonstrates strong separation performance and impressive zero-shot generalization ability using audio captions or text labels as queries, substantially outperforming previous audio-queried and language-queried sound separation models. Specifically, AudioSep achieved strong results including a Signal-to-Distortion Ratio Improvement (SDRi) of 7.74 dB across 527 sound classes of the AudioSet; 9.14 dB on the VGGSound dataset; 8.22 dB on the AudioCaps dataset; 6.85 dB on the Clotho dataset; 10.51 dB on the MUSIC dataset; 10.04 dB on the ESC-50 dataset; 8.16 dB on the DCASE 2024 Task 9 dataset; and an SSNR of 9.21 dB on the Voicebank-Demand dataset. For reproducibility of this work, we released the source code, evaluation benchmark and pre-trained model at: \url{https://github.com/Audio-AGI/AudioSep}. 
\end{abstract}


\section{Introduction}
\IEEEPARstart{C}{omputational} auditory scene analysis (CASA) \cite{wang2006computational} aims to design machine listening systems that perceive complex sound environments in a similar way to the human auditory system. As a fundamental research task for CASA, sound separation aims to separate real-world sound recordings into individual source tracks, also known as the “cocktail party problem" \cite{haykin2005cocktail}. Sound separation has a wide range of applications, including audio event separation \cite{lass, kavalerov2019universal}, music source separation \cite{kong2021decoupling}, and speech enhancement \cite{segan, kong2021speech}. 

Many previous works on sound separation mainly focus on separating one or a few sources, such as in speech enhancement \cite{segan, kong2021speech}, speech separation \cite{wang2018supervised, luo2019conv}, and music source separation \cite{kong2021decoupling}. Recently, universal sound separation (USS) \cite{kavalerov2019universal} has attracted many research interests. USS aims to separate arbitrary sounds in real-world sound recordings. Separating every sound source from a mixture is challenging due to the wide variety of sound sources existing in the world. As an alternative, query-based sound separation (QSS) has been proposed which aims to separate specific sound sources conditioned on a piece of query information. QSS allows users to extract desired audio sources which could be useful in many applications such as automatic audio editing \cite{rubin2013content} and multimedia content retrieval \cite{peng2017overview}. Using the query of different modalities such as vision \cite{SOP, chatterjee2021visual, audioscope}, audio \cite{kong2023universal, chen2022zero, gfeller2021one, wang2022few} or event labels \cite{veluri2023real, soundbeam, ochiai2020listen, wang2022improving} for sound separation has been investigated in the literature.

Recently, a new paradigm of QSS has been proposed, known as language-queried audio source separation (LASS) \cite{lass}. LASS is the task of separating arbitrary sound sources using natural language descriptions of the desired source. LASS provides a potentially useful tool for future digital audio applications, allowing users to extract desired audio sources via natural language instructions. The use of natural language queries offers significant advantages, compared to previous audio-visual \cite{SOP, chatterjee2021visual, audioscope} or audio-queried \cite{kong2023universal, chen2022zero, gfeller2021one, wang2022few} methods, such as flexible and convenient acquisition of query information. Compared to label-queried \cite{veluri2023real, soundbeam, ochiai2020listen, wang2022improving} methods that usually pre-define a fixed set of label categories, LASS does not limit the scope of input queries and can be seamlessly generalized to open domain.

The challenge of learning a LASS system is associated with the complexity and variability of natural language expressions. The text query description could range from sophisticated descriptions of multiple sound sources, such as “\textit{people speaking followed by music playing with a rhythmic beat}" to simple and compact phrases such as “\textit{speech, music}". In addition, the same audio source can be delivered with diverse language expressions, such
as “\textit{music is being played with a rhythmic beat}” or “\textit{an upbeat music melody is playing over and over again}”. LASS not only requires these phrases and their relationships to be captured in the language description but also that one or more sound sources that match the language query should be separated from the audio mixture. 

Our original approach \cite{lass} for LASS relies on supervised learning with labeled audio-text paired data. However, such annotated audio-text data is limited in size. To overcome the data scarcity issue, recent advancements have investigated training LASS with multimodal supervision \cite{clipsep, kilgour2022text, tan2023language}. The key idea behind this approach is to leverage multimodal contrastive pre-training models such as the contrastive language-image pretraining (CLIP) model, as the query encoder. As contrastive learning is capable of aligning text embedding with other modalities (e.g., vision), it enables the training of the LASS system using data-rich modalities and facilitates inference with text in a zero-shot mode. However, existing LASS methods leverage small-scale data for training and focus on separating restricted source types, such as musical instruments and a limited set of sound events. The potential to generalize LASS to open-domain scenarios, such as hundreds of real-world sound sources, has not yet been fully explored. Furthermore, previous works on LASS evaluate model performance on each domain-specific test set, which makes future comparison and reproduction inconvenient and inconsistent.

In this work, our goal is to establish a foundation model for sound separation with natural language descriptions. Our focus is on the development of a sound separation model, leveraging large-scale datasets, to enable robust generalization in open-domain scenarios. With this model, we aim to holistically address the separation of a diverse range of sound sources. This work constitutes an extension of our previous research, as presented in the Interspeech proceeding \cite{lass}. The contribution of this work includes:

\begin{itemize}
    \item We introduce AudioSep, a foundation model for open-domain, universal sound separation with natural language queries. AudioSep is trained with large-scale audio datasets and has shown strong separation performance and impressive zero-shot generalization capabilities. LASS-Net model presented in the Interspeech work did not perform well in open-domain scenarios.
    \item We extensively evaluate AudioSep. We construct a comprehensive evaluation benchmark for LASS research, involving numerous sound separation tasks such as audio event separation, musical instrument separation, and speech enhancement. We show that AudioSep substantially outperforms off-the-shelf audio-queried sound separation and state-of-the-art LASS models. AudioSep delivered strong results, such as an SDRi of \num{7.74} dB over \num{527} sound classes of the AudioSet; \num{9.14} dB on the VGGSound dataset; \num{8.22} dB on the AudioCaps dataset; \num{6.85} dB on the Clotho dataset; \num{10.51} dB on the MUSIC dataset; \num{10.04} dB on the ESC-50 dataset; \num{8.16} dB on the DCASE 2024 Task 9 dataset; and an SSNR of \num{9.21} dB on the Voicebank-Demand dataset.
    \item  We conduct in-depth ablation studies to investigate the impact of scaling up AudioSep using large-scale multimodal supervision \cite{clipsep, kilgour2022text, tan2023language}. 
    Our findings provide valuable insights for future research.
    \item We released the source code, evaluation benchmark, and pre-trained model at: \url{https://github.com/Audio-AGI/AudioSep} to promote research in this area. 
\end{itemize}

\section{Related Work}
\subsection{Audio Source separation}
Audio source separation~\cite{bell1995information} is a fundamental technique in signal processing, aimed at extracting independent source signals from their mixtures without prior knowledge of the mixing process. In recent years, audio source separation has shown significant advancements through integrating deep learning models~\cite{wang2018supervised}. Most deep learning-based source separation systems adopt the supervised learning approach, which usually requires the creation of simulated target and mixture sources and the developing of audio-to-audio mapping models. These mapping models generally fall into two categories: time domain-based and frequency domain-based approaches. Time domain-based methods focus on mapping directly onto the audio waveform, such as WaveUNet~\cite{stoller2018wave}, ConvTasNet~\cite{luo2019conv}, and Demucs~\cite{defossez2019demucs}. Frequency domain-based approaches conduct source separation within the spectral domain. While direct mapping on the audio spectrogram for audio source separation has demonstrated success~\cite{wang2020complex}, the mask-estimation-based methods, such as the estimation of ideal ratio masks~\cite{narayanan2013ideal}, is still the most widely used approach in the frequency domain source separation. Recent work has also explored the integration of both time and frequency supervisions~\cite{wan2023multi} to further enhance the performance and robustness systems. Besides the mapping-based approach, deep clustering-based methods~\cite{hershey2016deep, wang2018multi} have also shown promising results on speech separation based on the learning of a representation space with contrastive objectives.

\subsection{Universal sound separation}
A substantial amount of source separation research has been concentrated on domain-specific sound separation, focusing primarily on areas such as speech \cite{wang2018supervised, luo2019conv} or music \cite{kong2021decoupling}. Universal sound separation (USS) \cite{kavalerov2019universal} aims to separate a mixture of arbitrary sound sources in terms of their classes. The challenge inherent in USS is the diversity of sound classes in real-world scenarios, which increases the difficulty of separating all of these sound sources with a single sound separation system. The work in \cite{kavalerov2019universal} reported promising results on separating arbitrary sounds using permutation invariant training (PIT) \cite{PIT}, a supervised method initially designed for speech separation. The PIT method uses synthetic training mixtures simulated from single-source ground truth, performing sub-optimally due to a mismatch in the distribution between these synthetic mixtures and real-world sound recordings. Furthermore, it is not feasible to record a large database of single sources for PIT, as such sound recordings are often tainted by cross-talk. An unsupervised method called mixture invariant training (MixIT) \cite{MixIT} was proposed for sound separation using noisy audio mixtures. MixIT has achieved competitive performance compared to supervised methods (e.g., PIT), showing substantial improvements in reverberant sound separation performance. Both PIT and MixIT methods need a post-selection process to classify separated sources into specific sound classes.

\subsection{Query-based sound separation}
Query-based sound separation (QSS), also known as target source extraction, aims to separate a specific source from an audio mixture given some query information. Existing QSS approaches could be divided into three categories: audio-visual, audio-queried, and label-queried. 
\subsubsection{Audio-visual sound separation}
In the computer vision community, there has been active research focusing on utilizing visual information to extract target sounds in speech \cite{gao2021visualvoice, wu2019time}, music \cite{SOP}, and acoustic events \cite{chatterjee2021visual}. AudioScope \cite{audioscope} has been recently proposed to perform on-screen sound separation based on the MixIT method. Such vision-queried approaches are beneficial for automatically decomposing sound sources from audio-visual video data. However, their performance is affected by the dynamic visibility conditions of visual objects, and can be degraded in environments with visual occlusion or low-light conditions. In addition, video data often contains off-screen sounds. Learning from noisy video data is a key challenge for audio-visual sound separation systems \cite{audioscope, clipsep}.

\subsubsection{Audio-queried sound separation}
Another line of research leverages the audio modality as a query to separate acoustically similar sounds. Recent studies \cite{gfeller2021one, wang2022few, lee2019audio} have addressed the problem of using one or a few examples of a target source as the query to separate a particular sound source: this is known as one-shot or few-shot sound separation. These methods separate the target sound conditioned on the average audio embedding of a few audio examples of the target source, which requires labeled single sources for calculating query embedding during training. The work in \cite{kong2023universal, chen2022zero} proposes to train the audio-queried sound separation system with large-scale weakly labeled data (e.g., AudioSet \cite{audioset}), by first using a sound event detection model \cite{kong2020panns} to detect the anchor segment of sound events which are further used to constitute the artificial mixtures for the training of audio-queried sound separation models. These audio-queried sound separation approaches have shown great potential in the separation of unseen sound sources. However, during test time, the preparation of reference audio samples for the desired sound is often a time-consuming process.

\subsubsection{Label-queried sound separation} An intuitive way to query a specific sound source is to use the label of its sound class \cite{veluri2023real, soundbeam, ochiai2020listen, wang2022improving}. Although acquiring a label query involves less effort than acquiring an audio or visual query, the label set is often pre-defined and is limited to a finite set of source categories. This imposes a challenge when attempting to generalize the separation system into an open-domain scenario, which may require re-training the sound separation model, or the use of continual learning methods \cite{soundbeam, xiao2022continual}. Label lacks the capability to describe the relationship between multiple sound events, such as their spatial relation and temporal order.  This poses a challenge when the user intends to separate multiple sources rather than a single sound event.

\subsection{Language-queried audio source separation}
Language-queried audio source separation (LASS) is a recently proposed new paradigm of QSS. LASS uses a natural language description of an arbitrary target source to separate the source from an audio mixture. Such natural language descriptions can include auxiliary information for describing the target source, such as spatial and temporal relationships of sound events, for example “\textit{a dog barks in the background}” or “\textit{people applaud followed by a woman speaking}”. 

LASS-Net \cite{lass} was our first attempt to perform end-to-end language-queried sound separation. LASS-Net consists of a language query encoder and a separation model. The separation model performs target source separation in the frequency domain and the target waveform is reconstructed using the noisy phase and inverse short-time Fourier transform (iSTFT). LASS-Net is trained on a subset ($\sim$\num{17.3} hours, \num{33} sound categories) of the AudioCaps \cite{audiocaps} dataset and has shown great success in separating a wide range of sounds using audio caption queries. Kilgour et al. \cite{kilgour2022text} proposed a similar model that accepts audio or text queries in a hybrid manner. Tzinis et al. \cite{OCT} propose an optimal condition training (OCT) strategy for LASS. OCT performs greedy optimization toward the highest-performing condition among multiple conditions (e.g., signal energy, harmonicity) associated with a given target source, to improve the separation performance. 

These LASS methods \cite{lass, kilgour2022text, OCT} require labeled text-audio paired data for supervised training, while such labeled data is often limited in practice. Recent work \cite{clipsep, tan2023language} has investigated the potential of a self-supervised approach for LASS, including leveraging the visual modality as a bridge to learn the desired audio-textual correspondence. Instead of using a pre-trained language model (e.g., BERT \cite{bert}) \cite{lass}, Dong et al. \cite{clipsep} use the contrastive language-image pretraining (CLIP) \cite{clip} model as the query encoder and train a LASS model conditioned on the visual context of unlabeled noisy videos. Thanks to the aligned embedding space learned by the CLIP model, at the inference time, the separation model can be queried with text inputs in a zero-shot setting. Experimental results show that combining text and image conditions for hybrid training leads to better text-queried sound separation performance on musical instrument separation. Although preliminary studies have been undertaken into LASS, existing approaches work under the constraints of limited-source scenarios, such as musical instruments and a restricted set of universal sound classes. As such, these approaches have not met the expectation of LASS for zero-shot, open-domain sound separation.

\subsection{Multimodal audio-language learning}
Recently, the field of multi-modal audio-language has emerged as an important research area in audio signal processing and natural language processing. Audio-language tasks hold potential in various application scenarios. For instance, automatic audio captioning \cite{mei2021audio, drossos2017automated} aims to provide meaningful language descriptions of audio content, benefiting the hearing-impaired in comprehending environmental sounds. Language-based audio retrieval \cite{mei22_interspeech, oncescu2021audio} facilitates efficient multimedia content retrieval and sound analysis for security surveillance. Text-to-audio generation \cite{audioldm, ghosal2023text, huang2023make, yang2023diffsound, kreuk2022audiogen, liu2021conditional} aims to synthesize audio content based on language descriptions, serving as sound synthesis tools for film-making, game design, virtual reality, and digital media, and aiding text understanding for the visually impaired. Contrastive language-audio pre-training (CLAP) \cite{clap} aims to learn an aligned audio-text embedding space via contrastive learning. CLAP facilitates downstream audio-text multimodal tasks (e.g., zero-shot audio classification) \cite{liang2023adapting, audioldm}. In this work, we focus on the intersection between audio source separation and natural language processing, which is an important field for CASA research but less explored.

\section{Method}
\begin{figure*}
  \centering
  \includegraphics[width=\linewidth]{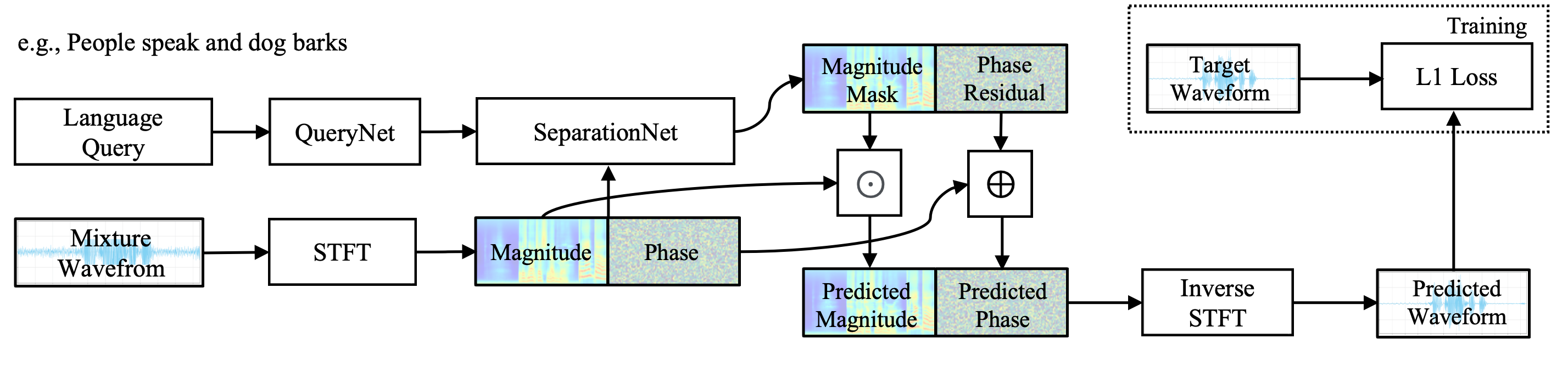}
  \caption{Framework of AudioSep. AudioSep has two key components: a QueryNet and a SeparationNet. The QueryNet is the text encoder of CLIP \cite{clip} or CLAP \cite{clap} model. The SeparationNet is a frequency-domain ResUNet \cite{kong2021decoupling, kong2023universal} model.}
  \label{fig-1}  
\end{figure*}

We introduce AudioSep, a foundation model for open-domain sound separation with natural language queries. AudioSep has two key components: a QueryNet and a SeparationNet, as illustrated in Figure \ref{fig-1}. We will introduce the details of each component below.

\subsection{QueryNet}
For QueryNet, we use the text encoder of the contrastive language-audio pre-training model (CLAP) \cite{clap}. The QueryNet is used to extract the text embedding of the natural language query. The input text query is denoted as $q=\{q_n\}_{n=1}^N$, consisting of a sequence of $N$ word tokens. The sequence is processed by a text encoder to obtain the text embedding for the input language query. The text encoder encodes the input text tokens via a stack of Transformer blocks. After passing through the transformer layers, the output representations are aggregated, resulting in a fixed-length $D$-dimensional vector representation, where $D$ corresponds to the latent dimension of the CLAP model. The text encoder is frozen during training.

\subsection{SeparationNet}
For SeparationNet, we apply the frequency-domain ResUNet model \cite{kong2021decoupling, kong2023universal} as the separation backbone. The input to the ResUNet model is a mixture of audio clips. First, we apply a short-time Fourier transform (STFT) on the waveform to extract the complex spectrogram $X \in \mathbb{C}^{T \times F}$, the magnitude spectrogram and phase of $X$ are denoted as $|X|$ and $e^{j{\angle}X}$, where $X = |X|e^{j{\angle}X}$. Then, we follow the same setting of \cite{kong2023universal}, and we construct an encoder-decoder network to process the magnitude spectrogram. The ResUNet encoder-decoder comprises \num{6} encoder blocks, \num{4} bottleneck blocks, and \num{6} decoder blocks. In each encoder block, the spectrogram is downsampled into a bottleneck feature using \num{4} residual convolutional blocks, while each decoder block utilizes \num{4} residual deconvolutional blocks to upsample the feature and obtain the separation components. A skip connection is established between each encoder block and the corresponding decoder block, operating at the same downsampling/upsampling rate. The residual block consists of \num{2} CNN layers, \num{2} batch normalization layers, and \num{2} Leaky-ReLU activation layers. Furthermore, we introduce an additional residual shortcut connecting the input and output of each residual block. The ResUNet model inputs the complex spectrogram $X$ and outputs the magnitude mask $|M|$ and the phase residual ${\angle}M$ conditioned on the text embedding $e_q$. $|M|$ controls how much the magnitude of $|X|$ should be scaled, and the angle $\angle{M}$ controls how much the angle of $\angle{X}$ should be rotated.
The separated complex spectrogram can be obtained by multiplying the STFT of the mixture and the predicted magnitude mask $|M|$ and phase residual $\angle{X}$:
\begin{equation}
    \hat{S} = |M| \odot |X|e^{j(\angle{X}+\angle{M})},
\end{equation}
where $\odot$ is the Hadamard product. The SeparationNet performs time-frequency masking in the magnitude domain, supplemented by phase correction relative to the noisy phase. An alternative approach is complex spectral mapping (CSM) \cite{wang2020complex}, which predicts the real and imaginary spectrograms jointly and has been shown to perform better than time-frequency masking in speech enhancement task. However, in this work, we focus on the time-frequency masking-based approach.

To bridge the text encoder and the separation model, we use a Feature-wise Linearly modulated (FiLm) layer \cite{perez2018film} after each ConvBlock deployed in the ResUNet. Specifically, let $H^{(l)} \in \mathbb{R}^{m \times h \times w}$ denote the output feature map produced by ConvBlock $l$ with $m$ channels, here $h$ and $w$ are the height and width of the feature map $H^{(l)}$, respectively. The modulation parameters are applied per feature map $H^{(l)}_i$ with the FiLm layer as follows:
\begin{equation}
    \label{eq-3}
    \operatorname{FiLM}(H^{(l)}_i|\gamma^{(l)}_i, \beta^{(l)}_i) = \gamma^{(l)}_iH^{(l)}_i + \beta^{(l)}_i
\end{equation}
where $H^{(l)}_i \in \mathbb{R}^{h \times w}$, and $\gamma^{(l)}, \beta^{(l)} \in \mathbb{R}^{m}$ are the modulation parameters from $g(.)$, i.e., $(\gamma, \beta)=g(e_q)$, such that $g(.)$ is a neural network and $e_q$ is the text embedding obtained from the text encoder. In this work, we model $g(\cdot)$ with two fully connected layers followed by ReLU activation, which is jointly trained with the ResUNet separation model.

\subsection{Loss and training}
During training, we use the loudness augmentation method proposed in \cite{kong2023universal}. When constituting the mixture \( x \) with \( s_1 \) and \( s_2 \), we first calculate the energy of \( s_1 \) and \( s_2 \) as \( E_1 \) and \( E_2 \) by \( E = \left\lVert s \right\lVert_2^2 \). We then determine a desired Signal-to-Noise Ratio (SNR) in decibels (dB), denoted as \(\text{SNR}_{\text{dB}}\), which is randomly chosen from the range \([-15, 15]\) dB. The scaling factor \(\alpha\) is then calculated using the formula:\[
\alpha = \sqrt{\frac{E_1}{E_2} \cdot 10^{\text{SNR}_{\text{dB}}/10}}\]
This scaling factor ensures that the mixture \( x \) has the specified SNR. The mixture \( x \) is then formed as follows:
\begin{equation}
    x = s_1 + \alpha s_2.
\end{equation}

We train AudioSep end-to-end using an L1 loss function between the predicted and target waveforms. Since waveform-based L1 loss is simple to implement and has shown good performance on universal sound separation tasks \cite{kong2023universal}.
\begin{equation}
    \operatorname{Loss}_{L1} = \left\lVert s-\hat{s} \right\lVert_1
\end{equation}
The lower L1 loss value indicates that the separated signal $\hat{s}$ is closer to the ground truth signal $s$. 

\subsection{Connections to the prior works}
The design of the AudioSep model is extended from our previous works, LASS-Net \cite{lass} and USS-ResUNet \cite{kong2023universal}. LASS-Net is a model for language-queried sound separation, while USS-ResNet is a model for audio-queried sound separation. The separation network, ResUNet, is the same across AudioSep, LASS-Net, and USS-ResNet. For the text encoder in AudioSep, we use CLAP \cite{clap}, which connects language and audio through an audio encoder and a text encoder via a contrastive learning objective. The CLAP model brings audio and text descriptions into a joint audio-text latent space, which facilitates effective audio-text representation learning.

\section{Datasets and Evaluation Benchmark}
In this section, we provide a detailed description of the training dataset used for AudioSep, along with the established evaluation benchmark. The statistics of the dataset are presented in Tables I and II.

\subsection{Training datasets}
\subsubsection{AudioSet} AudioSet \cite{audioset} is a large-scale, weakly-labelled audio dataset with \num{2} million \num{10}-second audio snippets sourced from YouTube. Each clip in the collection is categorised by the presented sound classes, without the timing information of sound events. AudioSet includes an ontology\footnote{\url{https://research.google.com/audioset/ontology/index.html}} of \num{527} distinct sound classes, such as ``Human sounds'', ``Music'', ``Natural Sounds'', among others. The training set comprises \num{2063839} clips, including a balanced subset of \num{22160} audio clips. For each sound class in the balanced subset, there are at least \num{50} audio clips. After accounting for unavailable YouTube links, we were able to download \num{1934187} audio clips with their corresponding video streams, equating to \num{94}\% of the complete training set. All the clips are either padded with silence or cut short to a duration of \num{10} seconds. Since a large amount of YouTube audio recordings have a sampling rate below \num{32} kHz, we have converted all audio recordings to a mono format and resampled them at \num{32} kHz. All audio clips within the training set are used as training data.

\subsubsection{VGGSound}
VGGSound \cite{vggsound} is a large-scale audio-visual dataset sourced from YouTube. VGGSound contains nearly \num{200000} video clips each of length \num{10} seconds, annotated across \num{309} sound classes consisting of human actions, sound-emitting objects, and human-object interactions. The creation process of VGGSound has ensured that the object producing each sound is also discernible in the corresponding video clip. We utilize the original version of the VGGSound dataset, which includes \num{183727} audio-visual clips for training and \num{15449} for testing. We resample all audio clips at \num{32} kHz. All data within the training split are used for training.

\subsubsection{AudioCaps}
AudioCaps \cite{audiocaps} is the largest publicly available audio captioning dataset, with \num{50725} \num{10}-second audio clips sourced from AudioSet. AudioCaps is divided into three splits: training, validation, and testing sets. The audio clips are annotated by humans with natural language descriptions through the Amazon Mechanical Turk crowd-sourced platform. Each audio clip in the training sets has a single human-annotated caption, while each clip in the validation and test set has five human-annotated captions. We retrieved AudioCaps based on the AudioSet we downloaded. The AudioCaps dataset we obtained consists of \num{49274} out of \num{49837} audio clips (\num{98}\%) in the training set, \num{494} out of \num{495} clips (\num{99}\%) in the validation set, and \num{957} out of \num{975} clips (\num{98}\%) in the test set. All audio clips within the training and validation sets are used for our training process.

\subsubsection{Clotho v2}
Clotho v2 \cite{clotho} is an audio captioning dataset that comprises sound clips obtained from the FreeSound platform\footnote{\url{https://freesound.org/}}. Each audio clip in Clotho has been human-annotated via the Amazon Mechanical Turk crowd-sourced platform. Particular attention was paid to fostering diversity in the captions during the annotation process. In this work, we use Clotho v2 which was released for Task \num{6} of the DCASE \num{2021} Challenge\footnote{\url{https://dcase.community/challenge2021}}. Clotho v2 contains \num{3839}, \num{1045}, and \num{1045} audio clips for the development, validation, and evaluation split respectively. The sampling rate of all audio clips in the Clotho dataset is \num{44100} Hz, each with five captions. Audio clips are of \num{15} to \num{30} seconds in duration and captions are \num{8} to \num{20} words long. We merge the development and validation split, forming a new training set with \num{4884} audio clips. All audio clips within the new training set and evaluation set are resampled at \num{32} kHz.

\subsubsection{WavCaps}
WavCaps \cite{wavcaps} is a recently released large-scale weakly-labeled audio captioning dataset, comprising \num{403050} audio clips with paired captions, totaling approximately \num{7568} hours. The audio clips constituting WavCaps originate from diverse sources, including FreeSound, BBC Sound Effects\footnote{\url{https://sound-effects.bbcrewind.co.uk/}}, SoundBible\footnote{\url{https://soundbible.com/}}, and AudioSet. The audio captions are filtered and generated using the assistance of ChatGPT\footnote{\url{https://openai.com/blog/chatgpt}} based on the online-harvested raw audio descriptions. The average duration of audio clips is \num{67.59} seconds and the average text length of captions is \num{7.8} words. We resampled all audio clips within WavCaps at \num{32} kHz for training.

\begin{table}[t]
\caption{AudioSep training datasets.}
\centering
\resizebox{0.95\columnwidth}{!}{
\begin{tabular}[\linewidth]{c c c c c c} 
 \hline
 & Caption & Label & Video & Num. clips & Hours \\ \hline
 AudioSet & $\times$ & \checkmark & \checkmark & \num{2063839} & \num{5800} \\
 VGGSound & $\times$ & \checkmark & \checkmark & \num{183727} & \num{550} \\
 AudioCaps & \checkmark & \checkmark & \checkmark & \num{49768} & \num{145} \\
 Clotho v2 & \checkmark & $\times$ & $\times$ & \num{4884} & \num{37}\\
 WavCaps & \checkmark & $\times$ & $\times$ & \num{40350} & \num{7568} \\
\hline
 
\end{tabular}}
\end{table}

\subsection{Evaluation benchmark}

\subsubsection{AudioSet}
The evaluation set of AudioSet \cite{audioset} contains \num{20317} audio clips with \num{527} sound classes. We downloaded
\num{18887} audio clips from the evaluation set (\num{93}\%) out of \num{20317} audio clips. Source separation on AudioSet presents a considerable challenge, given that AudioSet contains a diverse range of sounds within a hierarchical ontology. To create evaluation data, we adopt the pipeline proposed in \cite{kong2023universal}, which uses a sound event detection system \cite{kong2020panns} to analyze each \num{10}-second audio clip to pinpoint anchor segments. Subsequently, two anchor segments from different sound classes are selected and combined to form a mixture with a SNR of \num{0} dB. We generate \num{10} mixtures for each sound class, leading to \num{5270} mixtures for all \num{527} sound classes in total. 

\subsubsection{VGGSound}
In a similar way to \cite{clipsep}, we manually selected \num{100} clean samples that each contain a distinct target sound event from the VGGSound test set. We refer to this set of \num{100} samples as VGGSound-Clean. For each audio sample from VGGSound-Clean, we randomly selected 10 audio samples from the remaining VGGSound test set to generate mixtures. Specifically, following \cite{cosentino2020librimix}, we first uniformly sampled the loudness of the two audio samples between -\num{35} dB and -\num{25} dB LUFS (Loudness Units Full Scale); then we mixed the signals together. The mixtures were scaled to \num{0.9} if clipping occurred. Finally, we constructed an evaluation set with \num{1000} samples. The average SNR of the evaluation set is around \num{0} dB.

\subsubsection{AudioCaps}
Our downloaded test set of the AudioCaps dataset \cite{audiocaps} includes \num{957} audio clips, each annotated with five captions. To generate audio mixtures, we initially select an audio clip from the test set to serve as the target source, followed by a random selection of another audio clip as the background source, considering that the sound event tag\footnote{The sound event tags of each audio clip in AudioCaps can be retrieved from its corresponding AudioSet annotations.} of the background source does not coincide with that of the target source. For the test mixtures, each test audio is mixed with five randomly chosen background sources with an SNR at 0 dB. Each mixture is assigned one of the five audio captions of the target source. Consequently, \num{4785} test mixtures are created.

\subsubsection{Clotho v2}
The Clotho v2 \cite{clotho} evaluation set includes \num{1045} audio clips, each provided with five human-annotated captions. The duration of audio clips varies between \num{15} and \num{30} seconds. For the creation of test mixtures, we designate each audio clip in the evaluation set as a target source. Subsequently, we select two audio clips at random from the evaluation set, concatenate them, and then truncate it to match the length of the target source, thereby producing the interference source. Applying this pipeline, each audio clip in the evaluation set is mixed with five audio clips at an SNR of 0 dB. Each created mixture is then assigned one of the five audio captions of the target source. This procedure culminates in a total of \num{5225} mixtures for evaluation.

\subsubsection{ESC-50}
The ESC-50 dataset \cite{esc50} contains \num{2000} environmental audio recordings evenly arranged into \num{50} semantic classes including natural sounds, non-speech human sounds, domestic sounds, and urban noise. Each class contains \num{40} examples, with each audio clip having a duration of 5 seconds and with a sampling rate of \num{44.1} kHz. To make a consistent evaluation, we first downsample all audio clips at \num{32} kHz. Then, we randomly mix two audio clips from different sound classes with an SNR at \num{0} dB to form a pair. We constitute \num{40} mixtures for each sound class. This leads to a total of \num{2000} evaluation pairs, which are used to evaluate the zero-shot performance of our model on environmental sound separation with text label queries. 

\subsubsection{MUSIC}
The MUSIC dataset \cite{SOP} is a collection of \num{536} video recordings of people playing a musical instrument from \num{11} instrument classes such as accordion, acoustic guitar, and cello. These video clips are crawled from YouTube and are relatively clean. Following the previous work \cite{clipsep}, we downloaded the 46 video recordings in the test split. We further segmented all test videos into non-overlapping \num{10}-second clips and resampled them at \num{32} kHz. For each video segment, we randomly select one segment from each of the other instrument classes to create a mixture with an SNR at \num{0} dB, resulting in a total of \num{5004} evaluation pairs, which are used to evaluate the zero-shot performance of our model on musical instrument separation with text label queries.

\subsubsection{DCASE 2024 T9}
This is a new dataset\footnote{https://zenodo.org/records/10886481} collected for evaluating the performance of LASS systems for DCASE 2024 Task 9: Language-Queried Audio Source Separation\footnote{https://dcase.community/challenge2024/task-language-queried-audio-source-separation}. Specifically, \num{1000} audio files were sourced from the Freesound platform, uploaded between April and October 2023. Each audio file has been manually annotated with three captions. In the annotation guidance, annotators were instructed to describe the content of audio clips using five to twenty words. The tags of each audio file were verified and revised according to the FSD50K \cite{fsd50k} sound event categories. Each audio file has been chunked into a \num{10}-second clip and downsampled to \num{16} kHz. \num{3000} synthetic audio mixtures with signal-to-noise ratios (SNR) ranging from \num{-15}dB to \num{15}dB are generated for evaluation. The revised tag information is used to ensure that the two audio clips used in each mix do not share overlapping sound source classes. We use this dataset to evaluate the zero-shot performance of our model on universal sound separation with natural language queries. For models trained with \num{32} kHz audio, we first upsample all audio clips to \num{32} kHz for separation and then downsample the separated audio clips to \num{16} kHz for evaluation.

\subsubsection{Voicebank-DEMAND}
The Voicebank-DEMAND dataset \cite{voicebank} integrates the Voicebank dataset \cite{voicebank}, which includes clean speech, and the DEMAND \cite{DEMAND} dataset, which encompasses a variety of background sounds that are used to create noisy speech. The noisy utterances are created by mixing the Voicebank dataset and the DEMAND dataset under signal-to-noise ratios of \num{15}, \num{10}, \num{5}, and \num{0} dB. The test set of the Voicebank-DEMAND dataset includes a total of \num{824} utterances, which is used to evaluate the zero-shot performance of our model on speech enhancement. To make a fair comparison with previous speech enhancement systems \cite{segan, kong2021speech, scalart1996speech, macartney2018improved}, we resample all audio clips to \num{16} kHz. We use “Speech" as the input text query to perform speech enhancement.

\begin{table}[t]
\caption{The evaluation benchmark for open-domain sound separation with natural language queries.}
\centering
\resizebox{0.95\columnwidth}{!}{
\begin{tabular}[\linewidth]{c c c c c c c} 
 \hline
  & Num. mixtures & Sr (Hz) & Query Type \\ \hline
 AudioSet & \num{5270} & \num{32} k & text label & \\
 VGGSound & \num{1000} & \num{32} k & text label & \\
 AudioCaps & \num{4785} & \num{32} k & caption & \\
 Clotho v2 & \num{5225} & \num{32} k & caption & \\
 ESC-50 & \num{2000} & \num{32} k & text label & \\
 MUSIC & \num{5004} & \num{32} k & text label & \\
 DCASE 2024 T9 & \num{3000} & \num{16} k & caption & \\ 
 Voicebank-DEMAND & \num{824} & \num{16} k & text label & \\\hline
 
\end{tabular}}
\end{table}

\subsection{Evaluation metrics}
We utilize signal-to-distortion ratio improvement (SDRi) \cite{soundbeam, kong2023universal} and scale-invariant SDR (SI-SDR) \cite{sisdr} to evaluate the performance of sound separation systems. 
For the speech enhancement task, following previous works \cite{segan, kong2021speech, scalart1996speech, macartney2018improved}, we apply the perceptual evaluation of speech quality (PESQ) \cite{pesq}, mean opinion score (MOS) predictor of signal distortion (CSIG), MOS predictor of background-noise intrusiveness (CBAK), MOS predictor of overall signal quality (COVL) \cite{hu2007evaluation} and segmental signal-to-ratio noise (SSNR) \cite{quackenbush1988objective} for evaluation. For each evaluation metric, higher values indicate better performance.

\section{Experiments and Results}
\subsection{Training details}
We randomly sample two audio segments from two audio clips from the training set and mix them together to constitute a training mixture. The length of the audio segment is \num{5} seconds. We extract the complex spectrogram from the waveform signal with a Hann window size of \num{1024} and a hop size of \num{320}. 
For the CLAP model, we use the publicly-available state-of-the-art checkpoint `music\_speech\_audioset\_epoch\_15\_esc\_89.98.pt', which is trained on music, and speech datasets in addition to the original LAION-Audio-630k dataset \cite{clap}. 
For the separation model, we use a \num{30}-layer ResUNet consisting of \num{6} encoder and \num{6} decoder blocks, which is the same as the previous work \cite{kong2023universal} on universal sound separation. Each encoder block consists of two convolutional layers with kernel sizes of $3 \times 3$. The number of output feature maps of the encoder blocks is \num{32}, \num{64}, \num{128}, \num{256}, \num{512}, and \num{1024}, respectively. The decoder blocks are symmetric to the encoder blocks. We apply an Adam optimizer with a learning rate of \num{1e-3} to train the AudioSep with the batch size of \num{96}. We train the AudioSep model for \num{4}M steps on \num{8} Tesla V100 GPU cards.
\subsection{Comparison systems}

\subsubsection{LASS models}
We employ two state-of-the-art publicly available LASS models as the comparison systems. The first one is LASS-Net \cite{lass}, which uses a pre-trained BERT and ResUNet as the text query encoder and the separation model, respectively. LASS-Net is trained on a subset ($\sim$\num{17} hours) of AudioCaps including universal sounds of categories such as human sounds, animal, sounds of things, natural sounds, and environmental sounds. The second one is CLIPSep, which uses CLIP \cite{clip} as the query encoder and a model based on Sound-of-Pixels (SOP) \cite{SOP} for separation. CLIPSep \cite{clipsep} is trained with noisy audio-visual videos ($\sim$\num{500} hours) from VGGSound \cite{vggsound} dataset using hybrid vision and text supervision signal. In addition, CLIPSep employs a training strategy called noise invariant training (NIT), designed to enable robust learning from noisy video data. As the CLIPSep model is trained with \num{16} kHz sampling rate, for \num{32} kHz audio clips in the evaluation sets, we downsample the audio clips to \num{16} kHz for evaluation. Both LASS-Net and CLIPSep are frequency domain separation models, the noisy phase is used to reconstruct the waveform in the test time. Differences between the AudioSep and others in terms of the model size, architecture, training data are described in Table \ref{tab:model-compare}. AudioSep benefits from being trained on a significantly larger dataset, totaling \num{14100} hours, compared to the smaller datasets used by LASS-Net (\num{17.3} hours) and CLIPSep (\num{550} hours). We further show that the data scaling enables AudioSep to generalize better across various sound separation tasks.

\begin{table}[t]
\caption{Differences between the AudioSep and others in terms of the model size, architecture, training data.}
\centering
\resizebox{0.95\columnwidth}{!}{
\begin{tabular}[\linewidth]{c c c c c c } 
 \toprule
   & Training Data (hrs) & Parameters & Architecture \\ 
  \midrule
 LASSNet \cite{lass} & \num{17} & \num{63.4} M & BERT + ResUNet \\
 CLIPSep \cite{clipsep} & \num{550} & \num{181.6} M & CLIP + UNet \\
 USS-ResNet30 \cite{kong2023universal} & \num{5800} & \num{121} M  & CNN + ResUNet \\
 \midrule
 AudioSep & \num{14100} & \num{238.6} M   & CLAP + ResUNet \\
\bottomrule
\end{tabular}}
\label{tab:model-compare}
\end{table}

\subsubsection{Audio-queried sound separation models}
We use audio-queried separation models as comparison systems. Kong et al. \cite{kong2023universal} proposed a universal sound separation system trained on AudioSet ($\sim$\num{5800} hours). This system performs query-based sound separation by using the average audio embedding calculated from query examples from the training set as the condition and can separate hundreds of sound classes using a single model. Empirical studies \cite{kong2023universal} conducted by Kong et al. have assessed the effectiveness of various sound separation systems such as ConvTasNet \cite{luo2019conv}, UNet \cite{unet}, and ResUNet \cite{kong2021decoupling}, along with different audio embedding extractors such as PANNs \cite{kong2020panns} and HTSAT \cite{chen2022hts}. The results indicate that a system combining PANNs and ResUNet achieved the best performance. As a result, we adopt the PANN-ResUNet separation model with \num{30} and \num{60} layers of the ResUNet as the comparison systems, denoted as USS-ResUNet30 and USS-ResUNet60, respectively.

\subsubsection{Speech enhancement models}
Following \cite{kong2021speech}, we adopt four off-the-shelf speech enhancement models as the comparison systems including Wiener filter \cite{scalart1996speech}, SEGAN \cite{segan}, AudioSet-UNet \cite{kong2021speech}, Wave-U-Net \cite{macartney2018improved}, CCMGAN \cite{cmgan} and MP-SENet \cite{lu2023mp}. Wiener filter \cite{scalart1996speech} is a method based on signal processing methods. SEGAN is designed with the generative adversarial network \cite{goodfellow2020generative}. AudioSet-UNet is a frequency-domain UNet-based model trained with weakly labeled AudioSet \cite{audioset} data. Wave-U-Net is a time-domain UNet-based speech enhancement model. CMGAN \cite{cmgan} and MP-SENet \cite{lu2023mp} are two recently developed state-of-the-art models for speech enhancement. CMGAN \cite{cmgan} is a Conformer-based speech enhancement model optimized with MetricGAN \cite{fu2019metricgan}. MP-SENet \cite{lu2023mp} is an encoder-decoder architecture combining convolution-augmented transformers, designed to denoise magnitude and phase spectra simultaneously.

\begin{table*}
  \caption{Benchmark evaluation results of AudioSep and comparison with the state-of-the-art LASS systems.}
  \label{tab:main_results}
  \centering
  \resizebox{0.95\textwidth}{!}{
  \begin{tabular}{lccccccccccccccccc}
    \toprule
     & \multicolumn{2}{c}{\textbf{VGGSound}} & \multicolumn{2}{c}{\textbf{AudioCaps}} & \multicolumn{2}{c}{\textbf{Clotho}} & \multicolumn{2}{c}{\textbf{MUSIC}} & \multicolumn{2}{c}{\textbf{ESC-50}} & \multicolumn{2}{c}{\textbf{DCASE 2024 T9}} \\
	\cmidrule(lr){2-3} \cmidrule(lr){4-5} \cmidrule(lr){6-7}\cmidrule(lr){8-9}\cmidrule(lr){10-11}\cmidrule(lr){12-13}
     & SI-SDR & SDRi & SI-SDR & SDRi & SI-SDR & SDRi & SI-SDR & SDRi & SI-SDR & SDRi & SI-SDR & SDRi  \\
    \midrule
    LASSNet \cite{lass} & $-4.50$ & $1.17$ & $-0.96$ & $3.32$ & $-3.42$ & $2.24$ & $-13.55$ & $0.13$ & $-2.11$ & $3.69$ & $-4.01$ & $1.93$ \\
    CLIPSep \cite{clipsep} & $1.22$ & $3.18$ & $-0.09$ & $2.95$ & $-1.48$ & $2.36$ & $-0.37$ & $2.50$ & $-0.68$ & $2.64$ & $-1.09$ & $1.90$ \\
    \midrule
    AudioSep  & $\textbf{9.04}$ & $\textbf{9.14}$ & $\textbf{7.19}$ & $\textbf{8.22}$ & $\textbf{5.24}$ & $\textbf{6.85}$ & $\textbf{9.43}$ & $\textbf{10.51}$ & $\textbf{8.81}$ & $\textbf{10.04}$ & $\textbf{6.71}$ & $\textbf{8.16}$ \\
    \bottomrule
\end{tabular}}
\end{table*}

\subsection{Evaluation results on seen datasets}
We first assess the performance of AudioSep on datasets seen during training including AudioSet, VGGSound, AudioCaps, and Clotho, as shown in Table \ref{tab:main_results} and \ref{tab:as-results}. 
AudioSep shows strong sound separation performance using text labels or audio captions as input queries. 
On the AudioSet, AudioSep achieved an SI-SDR of \num{6.90} dB and an SDRi of \num{7.74} dB. For the VGGSound dataset, the AudioSep model demonstrated an SI-SDR of \num{9.04} dB and an SDRi of \num{9.14} dB. On the AudioCaps and Clotho datasets, the AudioSep achieved an SI-SDR of \num{7.19} dB and \num{5.24} dB, respectively, along with SDRi values of \num{8.22} dB and \num{6.85} dB, respectively.

Two state-of-the-art models, LASS-Net and CLIPSep, perform poorly in separating target sounds on our benchmarks. Audio-queried sound separation baseline systems USS-ResUNet30 and USS-ResUNet60 have achieved an SDRi of \num{5.57} dB and \num{5.7} dB, respectively, which underperformed AudioSep by around \num{2} dB. In the inference stage, AudioSep only requires the text description of the target source, which is more convenient to obtain, as compared with anchor audio required in the audio-queried methods. Evaluation results on seen datasets indicate the strong performance of AudioSep compared with previous state-of-the-art LASS models and off-the-shelf audio-queried sound separation models. 

\subsection{Zero-shot evaluation results}
We conducted further evaluations to assess the zero-shot separation performance of AudioSep on unseen datasets, including MUSIC, ESC-50, DCASE 2024 T9, and Voicebank-DEMAND. AudioSep achieved promising zero-shot separation results as shown in Table \ref{tab:main_results}. Specifically, for the MUSIC dataset, AudioSep achieved an SI-SDR of \num{9.43} dB and an SDRi of \num{10.51} dB. For the ESC-50 dataset, AudioSep obtained an SI-SDR of \num{8.81} dB and an SDRi of \num{10.04} dB. For the DCASE 2024 Task 9 dataset, AudioSep reached an SI-SDR of \num{6.71} dB and an SDRi of \num{8.16} dB. Although CLIPSep achieved good performance in separating musical instruments from background noise \cite{clipsep}, its performance degrades when faced with musical instrument mixtures. Both CLIPSep and LASS-Net perform poorly in these evaluation datasets.

Table \ref{tab:se-results} shows the results of Voicebank-DEMAND speech enhancement. Noisy speech without enhancement has PESQ, CSIG, CBAK, COVL, and SSNR of \num{1.97} dB, \num{3.35} dB, \num{2.44} dB, \num{2.63} dB and \num{1.68} dB respectively. AudioSep achieves PESQ, CSIG, CBAK, COVL, and SSNR of \num{2.43} dB, \num{3.37} dB, \num{3.17} dB, \num{2.88} dB and \num{9.21} dB respectively. AudioSep surpasses all the speech enhancement baselines in the PESQ metric and performs on par with traditional speech enhancement models including SEGAN \cite{segan} and Wave-U-Net \cite{macartney2018improved}. While the results achieved by AudioSep in speech enhancement are far from the state-of-the-art methods such as CMGAN \cite{cmgan} and MP-SENet \cite{lu2023mp}, this discrepancy may be attributed to the presence of large-scale noisy speech signals (e.g., AudioSet) in the training data. This noise can hinder the model's ability to effectively learn and separate clean speech signals, particularly in scenarios with an SNR range of \num{0} to \num{15} dB.

\begin{table}[t]
  \caption{Evaluation results of AudioSep and comparison with audio-queried sound separation systems on AudioSet.}
  \label{tab:as-results}
  \centering
  \resizebox{0.8\columnwidth}{!}{
  \begin{tabular}{lccc}
    \toprule
    & SI-SDR & SDRi & Query Modality \\
    \midrule
    USS-ResUNet\num{30} \cite{kong2023universal} & - & $5.57$ & Audio\\
    USS-ResUNet\num{60} \cite{kong2023universal} & - & $5.70$ & Audio\\
    \midrule
    AudioSep & $6.90$ & $7.74$ & Text\\
    \bottomrule
\end{tabular}}
\end{table}

\begin{table}[t]
\caption{Speech enhancement results on Voicebank-DEMAND dataset.}
\centering
\label{tab:se-results}
\resizebox{0.95\columnwidth}{!}{
\begin{tabular}[\linewidth]{c c c c c c c c c} 
    \toprule
  & PESQ & CSIG & CBAK & COVL & SSNR\\ 
      \midrule
 Noisy & $1.97$ & $3.35$ & $2.44$ & $2.63$ & $1.68$ \\
 Wiener \cite{scalart1996speech} & $2.22$ & $3.23$ & $2.68$ & $2.67$ & $5.07$ \\
 SEGAN \cite{segan} & $2.16$ & $3.48$ & $2.94$ & $2.80$ & $7.73$ \\
AudioSet-UNet \cite{kong2021speech} & $2.28$ & $2.43$ & $2.96$ & $2.30$ & $8.75$ \\
 Wave-U-Net \cite{macartney2018improved} & $2.40$ & $3.52$ & $3.24$ & $2.96$ & $9.97$ \\
CMGAN \cite{cmgan} & $3.41$ & $4.63$ & $3.94$ & $4.12$ & $11.10$ \\
MP-SENet \cite{lu2023mp} & $3.50$ & $4.73$ & $3.95$ & $4.22$ & $10.64$ \\
    \midrule
 AudioSep & $2.43$ & $3.37$ & $3.17$ & $2.88$ & $9.21$ \\
    \bottomrule
\end{tabular}}
\end{table}

\subsection{Visualization of separation results}
\begin{figure*}
  \centering
  \includegraphics[width=\linewidth]{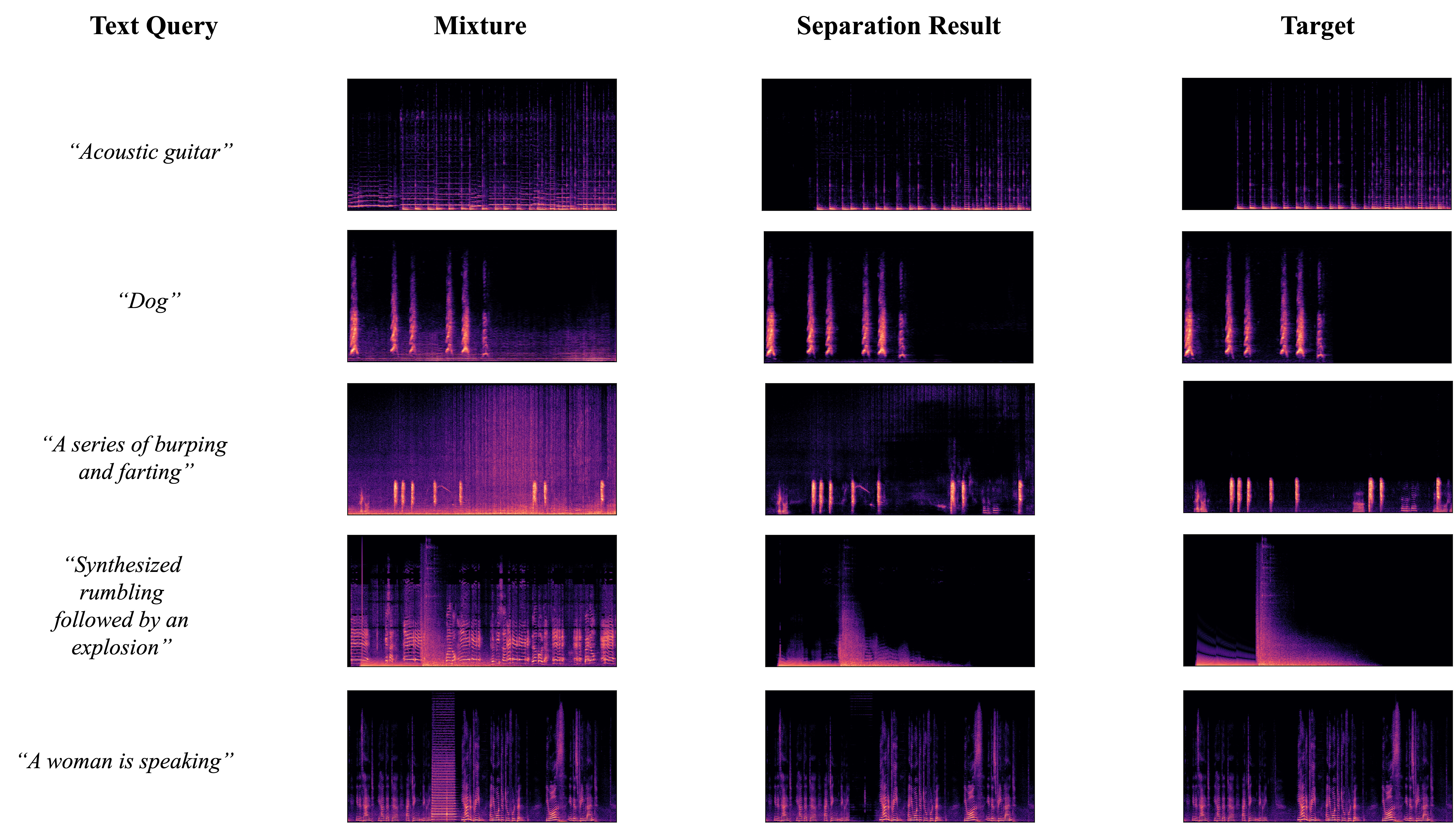}
  \caption{Visualization of separation results obtained by AudioSep.}
  \label{fig-2}  
\end{figure*}

We visualized spectrograms for audio mixtures, target audio sources, and separated sources using text queries of diverse synthetic mixtures (e.g., musical instruments, audio events, speech) with the AudioSep model, as shown in Figure \ref{fig-2}. The spectrogram patterns of the separated sources closely resemble those of the target sources, aligning with our objective experimental results. Additionally, we conducted several case studies using real audio clips sourcing from Freesound. The results demonstrated clear isolation of target audio components, highlighting the model's robustness and effectiveness. Audio samples are available on our project page\footnote{\url{https://audio-agi.github.io/Separate-Anything-You-Describe}}.

\section{Ablation Studies}

\subsection{Learning with Multimodal Supervision}
Recent research has explored the potential of using multimodal supervision \cite{clipsep, kilgour2022text, tan2023language} to enhance the scalability of training LASS models. For example, Contrastive Language-Image Pre-training (CLIP) model is pre-trained on large-scale image-text paired data using contrastive learning, where its text encoder learns to map textual descriptions into the same semantic space as visual representations. The key advantage of using the CLIP text encoder for LASS is that it enables us to train or scale up the LASS model using large-scale unlabeled audio-visual data \cite{audioset, vggsound}, leveraging visual embeddings as a substitute for annotated audio-text paired data. However, these works mainly focused on small-scale training sets (e.g., VGGSound \cite{vggsound}). In this section, we present ablation studies to investigate the efficacy of using large-scale multimodal supervision with CLIP and CLAP models to scale up AudioSep. We aim to gain insights into the applicability and performance of leveraging large-scale multimodal supervision for LASS.

\begin{table*}[t]
  \caption{Ablation study of scaling up AudioSep with multimodal supervision.}
  \vspace{1em}
  \label{tab:mm_studies}
  \centering
  \resizebox{\textwidth}{!}{
  \begin{tabular}{lccccccccccccccccccc}
    \toprule
    & \multicolumn{2}{c}{\textbf{AudioSet}} & \multicolumn{2}{c}{\textbf{VGGSound}} & \multicolumn{2}{c}{\textbf{AudioCaps}} & \multicolumn{2}{c}{\textbf{Clotho}} & \multicolumn{2}{c}{\textbf{MUSIC}} & \multicolumn{2}{c}{\textbf{ESC-50}} & \multicolumn{2}{c}{\textbf{DCASE 2024 T9}} \\
	\cmidrule(lr){2-3} \cmidrule(lr){4-5} \cmidrule(lr){6-7}\cmidrule(lr){8-9}\cmidrule(lr){10-11}\cmidrule(lr){12-13}\cmidrule(lr){14-15}
    & SI-SDR & SDRi & SI-SDR & SDRi & SI-SDR & SDRi & SI-SDR & SDRi & SI-SDR & SDRi & SI-SDR & SDRi & SI-SDR & SDRi  \\
    \midrule
    AudioSep-CLIP-TR0.0 & $0.45$ & $2.91$ & $-3.84$ & $0.78$ & $-0.02$ & $3.29$ & $-1.50$ & $2.42$ & $-0.21$ & $2.97$ & $-0.24$ & $3.67$ & $-2.49$ & $0.95$ \\
    AudioSep-CLIP-TR0.5 & $6.48$ & $7.28$ & $7.01$ & $7.27$ & $5.84$ & $7.25$ & $4.32$ & $6.10$ & $7.40$ & $9.05$ & $8.74$ & $9.97$ & $3.68$ & $5.18$ \\
    AudioSep-CLIP-TR0.75 & $6.52$ & $7.31$ & $\textbf{7.46}$ & $\textbf{7.67}$ & $5.98$ & $7.41$ & $4.38$ & $6.16$ & $8.00$ & $9.35$ & $8.92$ & $10.10$ & $3.94$ & $5.52$ \\
    AudioSep-CLIP-TR1.0 & $\textbf{6.60}$ & $\textbf{7.37}$ & $7.24$ & $7.50$ & $5.95$ & $7.45$ & $4.54$ & $6.28$ & $\textbf{9.14}$ & $\textbf{10.45}$ & $8.90$ & $10.03$ & $3.96$ & $5.46$ \\
    \midrule
    AudioSep-CLAP-TR0.0 & $-0.34$ & $3.32$ & $-1.64$ & $2.96$ & $2.12$ & $5.09$ & $0.63$ & $4.22$ & $4.42$ & $6.93$ & $1.24$ & $5.68$ & $0.65$ & $3.77$ \\
    AudioSep-CLAP-TR0.5 & $5.94$ & $6.88$ & $7.04$ & $7.24$ & $6.31$ & $7.62$ & $4.19$ & $6.13$ & $8.16$ & $9.65$ & $8.36$ & $9.63$ & $3.92$ & $5.38$ \\
    AudioSep-CLAP-TR0.75 & $5.94$ & $6.90$ & $7.20$ & $7.39$ & $6.28$ & $7.60$ & $4.29$ & $6.16$ & $8.42$ & $9.80$ & $8.83$ & $10.03$ & $3.65$ & $5.27$ \\
    AudioSep-CLAP-TR1.0 & $\textbf{6.58}$ & $\textbf{7.30}$ & $\textbf{7.38}$ & $\textbf{7.55}$ & $\textbf{6.45}$ & $\textbf{7.68}$ & $\textbf{4.84}$ & $\textbf{6.51}$ & $\textbf{8.45}$ & $\textbf{9.75}$ & $\textbf{9.16}$ & $\textbf{10.24}$ & $\textbf{4.47}$ & $\textbf{5.86}$  \\
    \bottomrule
\end{tabular}}
\end{table*}

\begin{table}[t]
\caption{Experimental results of comparison of different text queries on AudioCaps-Mini dataset. The suffix of TR1.0 is ignored.}
\label{tab:abs-various-text}
\centering
\resizebox{0.9\columnwidth}{!}{
\begin{tabular}[\linewidth]{c c c} 
 \toprule
  & SI-SDR & SDRi \\ 
  \midrule
 AudioSep-CLIP (text label) & \num{6.39} & \num{7.70} \\
 AudioSep-CLIP (original caption) & \num{7.27} & \num{8.25} \\
 AudioSep-CLIP (re-annotated caption) & \num{6.44} & \num{7.75} \\
  \midrule
 AudioSep-CLAP (text label) & \num{6.32} & \num{7.65} \\
 AudioSep-CLAP (original caption) & \num{7.73} & \num{8.47} \\
 AudioSep-CLAP (re-annotated caption) & \num{6.59} & \num{7.80} \\
 \bottomrule 
\end{tabular}}
\end{table}

\begin{table}[t]
\caption{An example from AudioCaps-Mini.}
\label{tab:example-audiocapsmini}
\centering
\resizebox{0.9\columnwidth}{!}{
\begin{tabular}[\linewidth]{c c } 
 \toprule
  & Text Description  \\ 
  \midrule
 Text label & “Vibration"  \\
 Original caption & “Clicking followed by vibrations"   \\
 Re-annotated caption1 & “Gear change, moving vehicle"   \\
 Re-annotated caption2 & “The distant engine sound"   \\
 Re-annotated caption3 & “The engine is running in distant"   \\
 Re-annotated caption4 & “The engine is starting"   \\
  \bottomrule
 
\end{tabular}}
\end{table}


Following \cite{clipsep}, we trained the proposed model using either CLIP or CLAP text encoders with a hyperparameter to control the training text rate (TR). The TR controls the percentage of training examples conditioned with text instead of audio or visual modality. For example, at \num{0}\% TR, the AudioSep model is trained exclusively with audio conditions from CLAP or visual conditions from CLIP. Conversely, at \num{100}\% TR, the model is trained solely with text conditions. Following \cite{clipsep}, we used the `ViT-B-32' checkpoint for the CLIP model. For video data processing, frames were uniformly extracted at one-second intervals, and their averaged CLIP embeddings were computed to serve as the query embeddings. Each model in this study was trained for \num{1}M steps on \num{8} Tesla V100 GPU cards. The model training applied a data augmentation method from \cite{kong2023universal}, which first augments the audio clips with the same energy level before mixing them together to create the mixture. 

For the CLIP-based models, we experiment with TRs of \num{0}\%, \num{50}\%, \num{75}\%, and \num{100}\%. The resulting models are referred to as AudioSep-CLIP-TR0.0, AudioSep-CLIP-TR0.5, AudioSep-CLIP-TR0.75, and AudioSep-CLIP-TR1.0, respectively. The visual condition is exclusively adopted for the training using AudioSet \cite{audioset} and VGGSound \cite{vggsound} datasets that have video stream information. For datasets that solely consist of audio and text, such as WavCaps \cite{wavcaps}, we use text condition. This training configuration allows us to leverage the available modalities of each dataset. Experimental results are shown in the upper part of Table \ref{tab:mm_studies}. When training AudioSep-CLIP-TR0.0 without text supervision from the AudioSet and VGGsound datasets, we observed clearly inadequate performance across all evaluation datasets. As large-scale video data may contain irrelevant audio contexts, this experimental result highlights the significance of text supervision from audio event labels provided by AudioSet and VGGSound in effectively training the LASS model. When training using text ratios of 50\% and 75\%, the overall performance is comparable to that of AudioSep-CLIP-TR1.0, which is trained exclusively with text supervision. This finding suggests that additional supervision from the visual modality does not improve separation performance in large-scale training settings, likely due to the inherent noise present in large-scale video data.

For the CLAP-based models, we use TRs of $0$\%, $50$\%, $75$\%, and $100$\%. The resulting models are denoted as AudioSep-CLAP-TR0.0, AudioSep-CLAP-TR0.5, AudioSep-CLAP-TR0.75, and AudioSep-CLAP-TR1.0. Experimental results are shown in the bottom part of Table \ref{tab:mm_studies}. At 0\% TR, where only audio modalities are used without text conditioning, the model underperforms, showing negative SI-SDR values on datasets like AudioSet and VGGSound, with only slightly better but still inadequate performance on others. For example, it achieves an SDRi of $4.22$ dB on Clotho and $6.93$ dB on MUSIC. This limited performance indicates that audio-only conditioning struggles in separating sources when text conditions are used during inference. However, as text supervision increases to 50\% and 75\% TR, the model’s performance improves substantially, indicating the important role of text conditioning during training. The separation performance continues to improve up to the TR1.0 setting, where the model is trained exclusively with text, achieving optimal results across all CLAP variants and demonstrating that text conditioning is crucial for effective representation learning in audio separation tasks. Overall, we observed that using additional audio supervision to scale AudioSep did not lead to improvements in our experiments, likely due to the \textit{modality gap} phenomenon \cite{liang2022mind}: the embedding spaces in the CLAP models may not be well aligned. A recent study \cite{saijo2024leveraging} demonstrated that it is possible to further improve the performance of AudioSep-CLAP model by including audio modality as supervision. However, this approach requires embedding manipulation techniques, such as applying dropout to the audio embeddings or adding Gaussian noise. While these techniques provide some improvement, the gains are not significant, we leave further exploration of this approach for future work.

Comparing the optimal results of CLIP-based and CLAP-based models, we observed that the CLAP-based model has better performance on audio datasets with natural language queries, such as AudioCaps, Clotho, and DCASE 2024 T9. This suggests that CLAP models may be more adept at learning representations from natural language queries. Both CLIP-based and CLAP-based models show comparable results on datasets where simple text labels serve as queries, including AudioSet, VGGSound, and ESC-50. This indicates that both models are equally effective when dealing with straightforward textual information (e.g., text labels). The CLIP model performs better on the MUSIC dataset, likely because its text embeddings capture distinctions between musical instruments more effectively than those of the CLAP model. Although CLIP was not trained on instrumental audio data, it was trained on a potentially vast amount of music-related image-text data\footnote{OpenAI has not released the training data for the CLIP model, so we are unable to verify its contents.}, enabling it to form clear distinctions in its text embedding space and achieve good performance in text-queried music instrument separation tasks. This observation is consistent with findings from the CLIPSep \cite{clipsep} experiments.

\begin{table}[h]
\caption{Examples of different invalid text queries.}
\label{tab-invalid}
\centering
\resizebox{0.9\columnwidth}{!}{
\begin{tabular}[\linewidth]{c c } 
 \toprule
  & Text Description  \\ 
  \midrule
 DIY  & “Separate Anything You Describe"   \\
 ESC-50  & “Church bells"   \\
 AudioCaps  & “Clicking followed by vibrations"   \\
 \midrule
 Ground truth  & “A man is laughing and then he says something."   \\
\bottomrule
\end{tabular}}
\end{table}

\begin{table}[h]
\caption{Experimental results of using different invalid text queries on DCASE 2024 T9 dataset.}
\label{tab:invalid}
\centering
\resizebox{0.6\columnwidth}{!}{
\begin{tabular}[\linewidth]{c c c} 
 \toprule
  Negative Caption & SI-SDR & SDRi \\ 
  \midrule
 DIY  & \num{-6.56} & \num{1.79} \\
 ESC-50 & \num{-7.68} & \num{1.48} \\
 AudioCaps  & \num{-8.51} & \num{1.35} \\
 \bottomrule 
\end{tabular}}
\end{table}

\begin{figure}[h]
  \centering
  \includegraphics[width=0.9\linewidth]{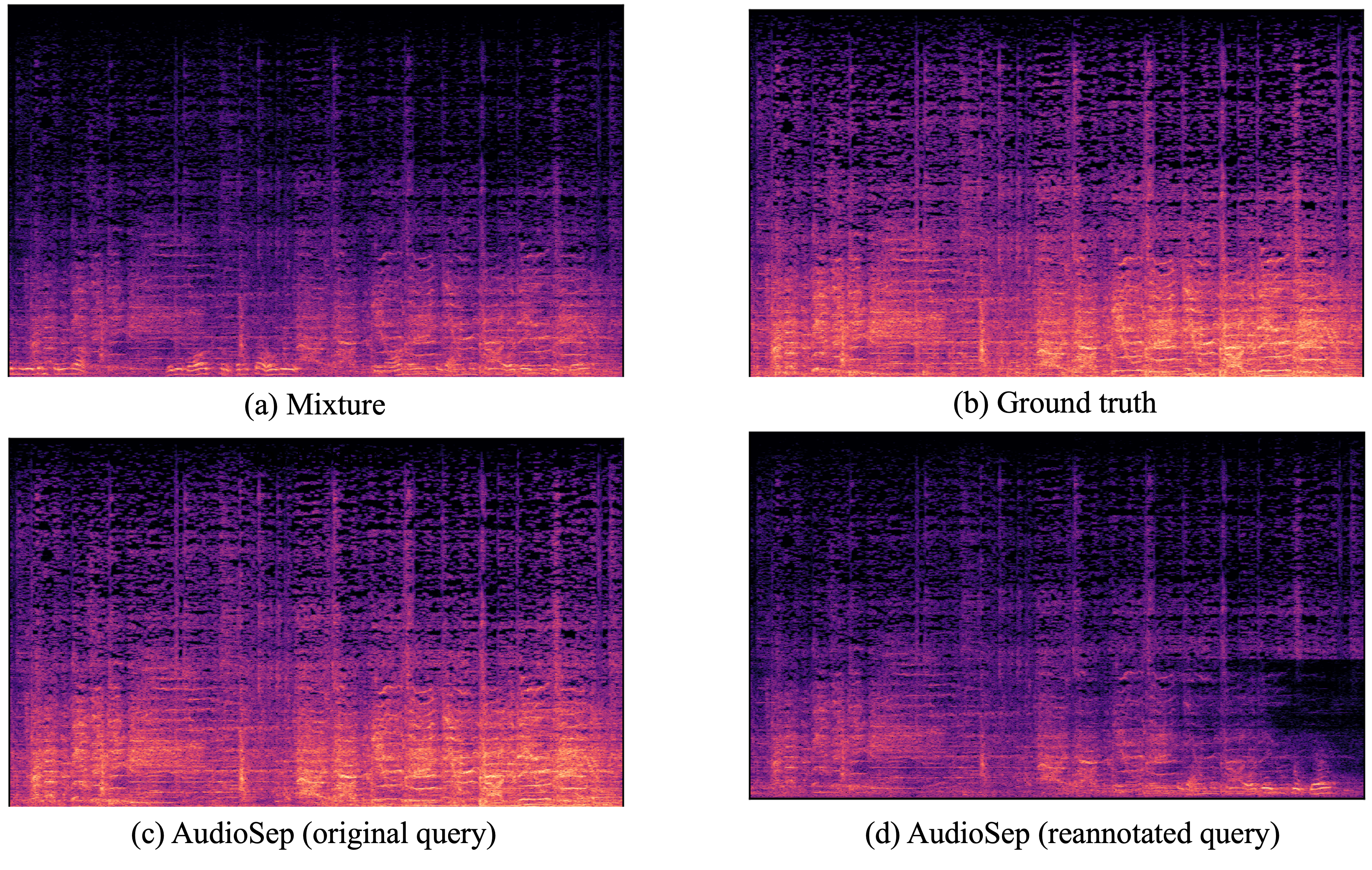}
  \caption{Case studies of (a) an audio mixture; (b) the ground truth target source; (c) the separated source queried by AudioCaps's original caption: \textit{“People laugh followed by people singing while music plays"}; (d) the separated source queried by our reannotated caption: \textit{“A music show is presenting to the public"}. Results are obtained using the AudioSep-CLAP-TR1.0 model.}
  \label{fig-various}  
\end{figure}

\begin{figure}[h]
  \centering
  \includegraphics[width=0.9\linewidth]{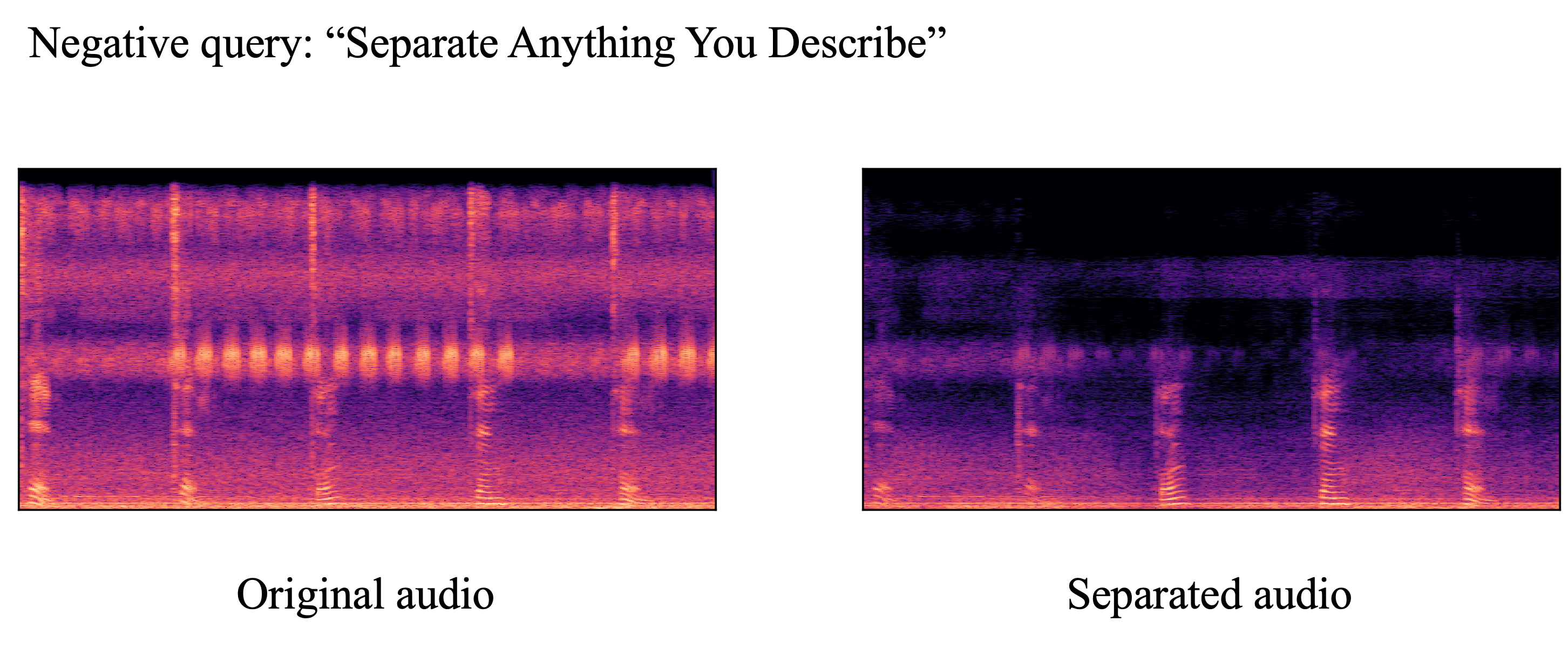}
  \caption{An example of performing separation using AudioSep with an invalid query: "Separate Anything You Describe".}
  \label{fig-neg}  
\end{figure}

\subsection{Studies of using various text queries}
In practice, human descriptions of an audio source are generally personalized, which poses a challenge to develop LASS models that can handle a variety of natural language queries well. In this section, we conduct an ablation study to investigate how does our proposed model perform when using various text queries. Specifically, we randomly selected \num{50} audio clips from the test set of the AudioCaps dataset and engaged four English native speakers from the University of Surrey to individually re-annotate the selected clips. Each annotator provided a single description per audio clip without any specific hints or restrictions. Consequently, we obtained a set of six descriptions for each audio clip. This set included one caption sourced from AudioCaps, four captions provided by the annotators, and the AudioSet audio event text labels. To create test mixtures, each audio clip was mixed with ten background sources randomly selected with an SNR at \num{0} dB, considering that the sound event labels of the background source do not overlap with that of the target source. We denote this test dataset as AudioCaps-Mini.

We evaluate the performance of AudioSep-CLIP-TR1.0 and AudioSep-CLAP-TR1.0 on AudioCaps-Mini. To investigate the effects of various text queries, we utilize the retrieved AudioSet event labels, the original AudioCaps audio captions, and our re-annotated natural language descriptions, which are referred to as the “text label" “original caption" and “re-annotated caption", respectively. An example of AudioCaps-Mini can be found in Table \ref{tab:example-audiocapsmini}. Experimental results are shown in Table \ref{tab:abs-various-text}. For both the CLIP-based and CLAP-based models, we have observed a considerable performance improvement when using the “original caption" as text queries instead of using the “text label". This may be attributed to the fact that human-annotated captions provide more comprehensive and accurate descriptions of the source of interest compared to audio event labels. Despite the personalized nature and different word distribution of our re-annotated captions, the results obtained using the “re-annotated caption" are slightly worse than those using the “original caption", while still marginally outperforming the results obtained with the “text label". These experimental findings demonstrate the promising generalization performance and robustness of AudioSep in real-world cases with diverse text queries.

We further investigated how the model's performance degrades when using reannotated queries compared to the original queries. We observed several instances where the model's performance degrades slightly as distortion increases. An example could be found in Figure \ref{fig-various}.

\subsection{Studies of using invalid text queries}
In this section, we perform an ablation study to investigate the effect of using invalid queries, i.e., queries that do not match any sources presented in the audio mixtures, on the performance of audio source separation. We used three different types of invalid queries: one randomly sampled caption from the AudioCaps test set, another from the ESC-50 event class labels, and a DIY text query labeled “Separate Anything You Describe." Examples of these different invalid queries are presented in Table \ref{tab-invalid}. We conducted this ablation study on DCASE 2024 Task 9 evaluation set and using the pre-trained AudioSep model.

The separation results are presented in Table \ref{tab:invalid}. For all three types of invalid queries, the SI-SDR values are negative. Through manual inspection, we observed that the AudioSep model tends to suppress some signals randomly when presented with invalid queries. An example can be found in Figure \ref{fig-neg}. As potential optimizations, we could consider supervising the model to generate silence in response to invalid queries or leveraging novel methods proposed in the OCT \cite{OCT}. We leave these for our future work.

\section{Conclusion and Future Work}
We have introduced AudioSep, a foundation model for open-domain universal sound separation with natural language descriptions. AudioSep can perform zero-shot separation using text labels or audio captions as queries. We have presented a comprehensive evaluation benchmark including numerous sound separation tasks such as audio event separation, musical instrument separation, and speech enhancement. AudioSep outperforms state-of-the-art text-queried separation systems and off-the-shelf audio-queried sound separation models. We show that AudioSep is a promising approach to flexibly address the CASA problem with strong sound separation performance. In future work, we will improve the separation performance of AudioSep via unsupervised learning techniques \cite{audioscope, MixIT}  and extend AudioSep to support audio-visual sound separation, audio-queried sound separation, and text-guided speaker separation \cite{hao2023typing} tasks.

\section*{Acknowledgments}
This work is partly supported by UK Engineering and Physical Sciences Research Council (EPSRC) Grant EP/T019751/1 ``AI for Sound'', British Broadcasting Corporation Research and Development~(BBC R\&D), and a PhD scholarship from the Centre for Vision, Speech and Signal Processing, Faculty of Engineering and Physical Science, University of Surrey. For the purpose of open access, the authors have applied a Creative Commons Attribution (CC-BY) license to any Author Accepted Manuscript version arising. We thank the reviewers and associate editor for their constructive comments to further improve the paper.

\bibliography{reference}
\bibliographystyle{IEEEtran}

\vfill

\end{document}